\documentclass[]{aa}  
\usepackage{graphicx}
\usepackage{subfig}

\usepackage[usenames,dvipsnames]{xcolor}
\usepackage[varg]{txfonts}
\usepackage{longtable}
\usepackage{threeparttable}

\def\intd{\,\mathrm{d}}

\usepackage{soul}

\begin{document} 

\title{A panoptic model for planetesimal formation and pebble delivery}

   \author{S.~Krijt\inst{1}\thanks{Now at the Department of the Geophysical Sciences, The University of Chicago, 5734 South Ellis Avenue, Chicago, IL 60637, USA; \email{skrijt@uchicago.edu}}
          \and
          C.\,W.~Ormel\inst{2}
          \and
          C.~Dominik\inst{2}
          \and
          A.\,G.\,G.\,M.~Tielens\inst{1}}
      
\institute{Leiden Observatory, Leiden University, Niels Bohrweg 2, 2333 CA Leiden, The Netherlands\label{inst1}
\and Anton Pannekoek Institute, University of Amsterdam, Science Park 904, 1098 XH Amsterdam, The Netherlands\label{inst2}}

   \date{}

 \abstract
{The journey from dust particle to planetesimal involves physical processes acting on scales ranging from micrometers (the sticking and restructuring of aggregates) to hundreds of astronomical units (the size of the turbulent protoplanetary nebula). Considering these processes simultaneously is essential when studying planetesimal formation.}
 {The goal of this work is to quantify where and when planetesimal formation can occur as the result of porous coagulation of icy grains and to understand how the process is influenced by the properties of the protoplanetary disk.}
  {We develop a novel, global, semi-analytical model for the evolution of the mass-dominating dust particles in a turbulent protoplanetary disk that takes into account the evolution of the dust surface density while preserving the essential characteristics of the porous coagulation process. This panoptic model is used to study the growth from submicron to planetesimal sizes in disks around Sun-like stars.}
 {For highly porous ices, unaffected by collisional fragmentation and erosion, rapid growth to planetesimal sizes is possible in a zone stretching out to ${\sim}10\mathrm{~AU}$ for massive disks. When porous coagulation is limited by erosive collisions, the formation of planetesimals through direct coagulation is not possible, but the creation of a large population of aggregates with Stokes numbers close to unity might trigger the streaming instability (SI). However, we find that reaching conditions necessary for SI is difficult and limited to dust-rich disks, (very) cold disks, or disks with weak turbulence.}
 {Behind the snow-line, porosity-driven aggregation of icy grains results in rapid (${\sim}10^{4}\mathrm{~yr}$) formation of planetesimals. If erosive collisions prevent this, SI might be triggered for specific disk conditions. The numerical approach introduced in this work is ideally suited for studying planetesimal formation and pebble delivery simultaneously and will help build a coherent picture of the start of the planet formation process.}

\maketitle

\section{Introduction}
Protoplanetary disks are the sites of planet formation. In these disks, the coagulation of microscopic dust particles, already present in the interstellar medium, into kilometer-sized planetesimals constitutes the first -- and arguably least understood -- step in the assembly of fully grown planets \citep{testi2014,johansen2014}. Initially, the dust aggregates, held together by surface forces, grow by sticking to each other in gentle, low-velocity collisions \citep[e.g.,][]{kempf1999}. As a result, aggregates form very open, porous structures. As the aggregates gain mass and their relative velocities increase, collisions become more energetic, possibly leading to compaction and catastrophic fragmentation \citep{dominiktielens1997,blumwurm2000,ormel2007}. A second hurdle is presented in the form of the radial drift barrier: particles with certain aerodynamic properties decouple from the gas and drift radially on short timescales \citep{whipple1972, weidenschilling1977,brauer2007}.

In the inner few AU of the protoplanetary disk, dust grains consist mainly of silicates, and these aggregates bounce off each other in collisions or even disrupt completely upon impact at collision velocities above several $\mathrm{m~s^{-1}}$ \citep{blumwurm2008,guttler2010}. These collisional processes limit the growth beyond a centimeter or so in the inner disk \citep{brauer2007, zsom2010, windmark2012}.

Outside the snow line, located typically at ${\sim}3\mathrm{~AU}$ \citep{min2011}, water ice becomes an important constituent of the dust grains. This is beneficial for growth because aggregates composed mostly of ice are capable of sticking at tens of $\mathrm{m~s^{-1}}$ \citep{wada2009,wada2013,gundlach2014}. In addition, these icy particles maintain highly porous structures \citep{suyama2008,suyama2012}, making them less likely to bounce in collisions \citep{wada2011, seizinger2013a} and allowing them to out-grow the radial drift barrier in the inner ${\sim}10\mathrm{~AU}$ of the protoplanetary nebula \citep{okuzumi2012,kataoka2013c}. However, the growth of these porous aggregates might be frustrated by high-velocity erosive collisions \citep{krijt2014b} or sintering in certain regions of the disk \citep{sirono2011}.

Instead of coagulating directly, planetesimals can also be formed through particle concentration mechanisms \citep[][and references therein]{johansen2014}. One promising mechanism is the streaming instability (SI) \citep{youdin2005,johansen2007N,bai2010a,bai2010b}, which can be triggered by a dense midplane layer of partially decoupled dust particles. Recently, \citet{drazkowska2014b} have defined a set of conditions for SI and compared them to dedicated models of compact coagulation. They found that in the inner disk, where the growth of silicates is limited by bouncing/fragmentation, particles cannot grow to Stokes numbers\footnote{A measure for the degree of coupling between the dust particle and the surrounding gas.} large enough ($\mathrm{St}\sim10^{-2}-1$) for triggering SI. Outside the snow-line however, rapidly growing highly porous ice aggregates can grow to large Stokes numbers \citep{okuzumi2012} at which point their growth is possibly limited by erosive collisions \citep{krijt2014b}. The possibility of triggering SI through rapid porous coagulation has not yet been investigated, but it has been shown that erosion-limited porous growth can concentrate most of the dust mass in $\mathrm{St}\sim1$ particles \citep{krijt2014b}.

We set out to study the formation of the first generation of planetesimals. Giant planets have not yet formed, hence the protoplanetary disk is smooth. The focus is to understand the evolution of the mass-dominating particles in disks around Sun-like stars and understand how their evolution influences the dust surface density. Ultimately, the goal is to identify regions in both space and time where the first planetesimals can form, either through direct porous coagulation \citep[e.g.,][]{okuzumi2012} or through coagulation triggering SI \citep{drazkowska2014b}. 

In order to answer these questions, we develop a global, semi-analytical, panoptic model that captures the evolution of the mass-dominating particles as they grow and drift radially in the protoplanetary disk (Sect. \ref{sec:description}), including a detailed description of grain porosity (Sect. \ref{sec:include_porosity}) and erosion (Sect. \ref{sec:include_erosion}). After testing the method against two well-studied cases in Sect. \ref{sec:test_cases}, we use it to study rapid porous growth through the drift barrier in Sect. \ref{sec:porous} and erosion-limited porous growth as a possible cause for SI (Sect. \ref{sec:porous_erosion}). The results are discussed in Sect. \ref{sec:discussion} and conclusions are presented in Sect. \ref{sec:conclusions}.

\section{Nebula model and dust properties}\label{sec:method}
We consider a turbulent disk of gas around a $1M_\odot$ star. Neglecting the possible presence of pressure bumps, dead zones, etc., we concentrate on the outer regions where ices are an important part of the solid mass reservoir.

\subsection{Nebula model}\label{sec:nebula}
We adopt a truncated power-law for the gas surface density distribution,
\begin{equation}\label{eq:Sigma_g}
\Sigma_\mathrm{g} = \begin{cases} 
~\Sigma_\mathrm{g,0} r^{-\gamma} & \textrm{~~for $r\leq r_\mathrm{out}$}, \vspace{3mm} \\
~0 & \textrm{~~for $r>r_\mathrm{out}$}. 
\end{cases}
\end{equation}
The normalization constant will be determined by fixing the total mass of the gas disk $M_\mathrm{D}$, using
\begin{equation}
\Sigma_\mathrm{g,0} =  (2-\gamma) \frac{ M_\mathrm{D}}{2\pi  r_\mathrm{out}^{2-\gamma}}.
\end{equation}
Typical power law exponents range between $\gamma=3/2$, appropriate for the Minimum Mass Solar Nebula (MMSN) \citep{hayashi1981}; and $\gamma \simeq 1$, consistent with observations \citep{andrews2009} and accretion disk theory \citep[e.g.,][]{armitage2010}. Disk masses between $M_\mathrm{D} = 10^{-3}M_\odot$ and $0.2M_\odot$ and radii between $30-100\mathrm{~AU}$ are consistent with observational constraints for disks in the Taurus star forming region \citep{andrews2005,andrews2013}, though the drop in surface density in the outer edge is found to be much less dramatic than the drop-off in Eq. \ref{eq:Sigma_g}.

Irrespective of the disk mass, we adopt a temperature structure
\begin{equation}\label{eq:T}
T = 125 \left(\frac{r}{5\mathrm{~AU}}\right)^{-1/2} \mathrm{~K},
\end{equation}
appropriate for an optically thin disk and in agreement with observational constraints \citep{andrews2005}. 

Most other quantities are derived, together with assumptions about the turbulence and vertical structure, from Eqs. \ref{eq:Sigma_g} and \ref{eq:T}. The isothermal gas sound speed is given by
\begin{equation}
c_\mathrm{s} = \sqrt{k_{\mathrm{B}} T / m_\mathrm{g}},
\end{equation}
with $k_{\mathrm B}$ Boltzmann's constant and $m_\mathrm{g} = 2.34 m_\mathrm{p} = 3.9\times10^{-24}\mathrm{~g}$ with $m_\mathrm{p}$ the mass of a proton. The Kepler frequency equals
\begin{equation}
\Omega = \sqrt{GM_{\odot}/r^3} = 1.8\times10^{-8} \left(\frac{r}{5\mathrm{~AU}}\right)^{-3/2} \mathrm{~s^{-1}}.
\end{equation} 
Assuming an isothermal column, the gas density drops with increasing distance from the midplane $z$ according to
\begin{equation}\label{eq:rho_g}
\rho_\mathrm{g} = \frac{\Sigma_\mathrm{g}}{\sqrt{2\pi}h_\mathrm{g}}\exp\left(\frac{-z^2}{2h_{\mathrm{g}}^2} \right),
\end{equation}
with the vertical scale height of the gas $h_\mathrm{g} = c_\mathrm{s} / \Omega$. The turbulent viscosity is parametrized as $\nu_\mathrm{turb} = \alpha c_\mathrm{s}^2  / \Omega$ following \citet{shakura1973}. Unless noted otherwise, we will use the disk parameters listed in Table \ref{tab:benchmark}.

\begin{table}
\centering
\begin{tabular}{l c c}
\hline\hline
Quantity & symbol & value \\
\hline
Total disk mass & $M_\mathrm{D}$ & $10^{-2}M_\odot$ \\
Disk cut-off radius & $r_\mathrm{out}$ & $100\mathrm{~AU}$\\
Surface density exponent & $\gamma$ & 1.5\\
Initial dust-to-gas ratio & $Z_0$ & $0.02$ \\
Turbulence strength & $\alpha$ & $10^{-3}$ \\
 \hline\hline
\end{tabular}
\caption{The benchmark disk model used throughout this work.}
\label{tab:benchmark}      
\end{table}

\subsection{The dust content}\label{sec:batch}
Initially, the dust density follows the gas density through $\Sigma_\mathrm{d}/\Sigma_\mathrm{g} = Z_0$, with $Z_0=0.02$ the vertically integrated metallicity. The individual dust particles are assumed to be compact spherical monomers with radii $a_\bullet$ and masses $m=m_\bullet = (4/3)\pi \rho_\bullet a_\bullet^3$. We use $\rho_\bullet=1.4\mathrm{~g~cm^{-3}}$, appropriate for icy particles. When dust particles grow, they are uniquely described by a mass $m$ and volumetric filling factor $\phi$ \citep[e.g.,][]{krijt2014b}. Depending on the details of the growth process (i.e., the disk properties and collision physics), the filling factor (a measure for the porosity) can be as low as ${\sim}10^{-5}$ or very close to 1, resembling compact spheres.

\subsubsection{Vertical and radial motions}

The relative dust scale height is given by \citep{youdin2007}
\begin{equation}\label{eq:h_d}
\frac{h_\mathrm{d}}{h_\mathrm{g}} = \left(1+ \frac{\Omega t_s}{\alpha}\frac{1+2\Omega t_s}{1+\Omega t_s} \right)^{-1/2}.
\end{equation}
The dust scale height depends on the particle Stokes number $\Omega t_s$, with the stopping time $t_s$ a function of the particle mass and porosity, and the local gas properties. Appendix \ref{sec:t_s} describes how the stopping time can be calculated.

The radial drift velocity is then given by \citep{weidenschilling1977}
\begin{equation}\label{eq:v_drift}
\dot{r} = - v_\mathrm{drift} = - \frac{2\Omega t_s}{1+(\Omega t_s)^2} \eta v_\mathrm{K},
\end{equation}
where $v_\mathrm{K}=R \Omega$ is the Keplerian orbital velocity and $\eta$ can be written as \citep{nakagawa1986}
\begin{equation}\label{eq:eta}
\eta \equiv - \frac{1}{2} \left( \frac{c_\mathrm{s}}{v_\mathrm{K}} \right)^2 \frac{\partial \ln(\rho_\mathrm{g} c_\mathrm{s}^2)}{\partial \ln r} \approx \left( \frac{c_\mathrm{s}}{v_\mathrm{K}} \right)^{2}.
\end{equation}
The drift timescale is defined as 
\begin{equation}\label{eq:t_drift}
t_\mathrm{drift}\equiv - r / \dot{r}
\end{equation}
and depends on the masses and porosities of the dust particles through their dimensionless stopping time $\Omega t_s$ (see Appendix \ref{sec:t_s}). Equation \ref{eq:v_drift} neglects the backreaction of the dust particles on the gas and overestimates the drift velocity when the dust density becomes comparable to the gas density. In most models presented here such conditions are not reached and identifying regions and times where such collective effects become important is one of the goals of this paper.

\subsubsection{Growth timescales in the single-size approximation}
The model for porous coagulation is based on the semi-analytical model of \citet[][Section 5]{krijt2014b}. At the heart of the semi-analytical approach lies the assumption that the local dust population can be approximated by a mono-disperse grain population, with a single characteristic mass and characteristic porosity. This assumption is valid when \emph{i)} the full mass distribution has a clearly defined peak mass and porosity; and \emph{ii)} the growth (and porosity-evolution, if dominated by collisions) of the peak-mass grains is mainly due to collisions with similar-size particles. These assumptions generally hold for populations resulting from porous coagulation \citep[e.g.,][]{ormel2007,okuzumi2012}.

\begin{figure}
\centering
\includegraphics[clip=,width=.95\linewidth]{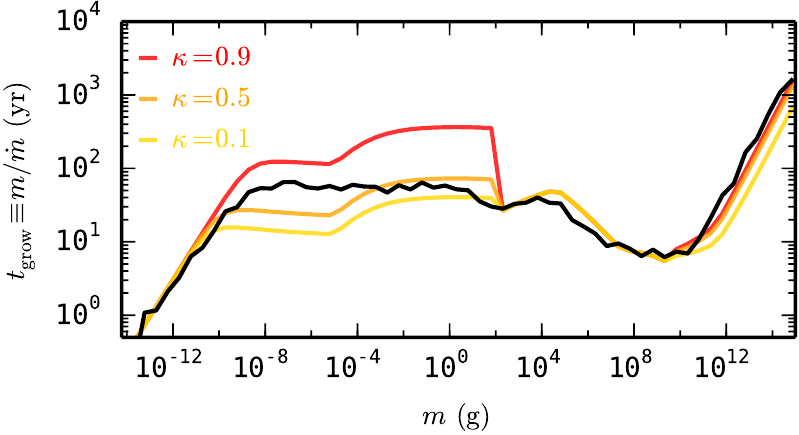}
\caption{Growth timescales of highly porous aggregates in an MMSN-like disk at 5 AU with a turbulence of $\alpha=10^{-3}$. The solid black line is obtained with Monte Carlo models that take into account the full particle size and porosity distributions  \citep{krijt2014b} and the colored lines show Eq. \ref{eq:m_dot} for three different values of $\kappa$ (see text).}
\label{fig:t_grow}
\end{figure}

For a mono-disperse dust population, and assuming all collisions result in perfect sticking, the growth rate is given by
\begin{equation}\label{eq:m_dot}
\dot{m} = \frac{\Sigma_\mathrm{d}}{\sqrt{2\pi} h_\mathrm{d}} \sigma_\mathrm{col} v_\mathrm{rel},
\end{equation}
with $h_\mathrm{d}$ the dust scale height, $\sigma_\mathrm{col}$ the collisional cross section, and $v_\mathrm{rel}$ the relative velocity between the (identical) dust particles. The growth timescale is then defined as
\begin{equation}\label{eq:t_grow}
t_\mathrm{grow} \equiv m / \dot{m}.
\end{equation}
Appendix \ref{sec:v_rel} describes how to calculate $v_\mathrm{rel}$ between two grains with arbitrary properties.

In reality however, the dust size distribution will not be infinitely narrow and dust grains will grow by colliding with dust particles with a broad range of masses and porosities, see for example \citet[][Fig. 9]{okuzumi2012} and \citet[][Fig. 10]{krijt2014b}. In order to simulate the width of the distribution, instead of calculating the velocity between \emph{identical} particles, we calculate the velocity between two particles with Stokes numbers $\Omega t_s$ and $\kappa \Omega t_s$, with $0 \leq \kappa \leq 1$ a numerical factor\footnote{See \citet{sato2015} for further discussion on this approach in the case of compact growth.}. The two limiting cases refer to the relative velocity between identical particles ($\kappa=1$) and between a particle and an extremely well-coupled small grain ($\kappa = 0$). 

In Fig. \ref{fig:t_grow}, we compare Eq. \ref{eq:m_dot} to the full Monte Carlo model of \citet{krijt2014b}. For this comparison, radial drift is neglected and perfect sticking is assumed. For every particle mass, the corresponding porosity is calculated by assuming an initial hit-and-stick phase that is followed by gas-compaction and eventually self-gravity compaction (see Sect. \ref{sec:include_porosity} and Appendix \ref{sec:phi} for more details). The disk model is identical to the one in \citet{okuzumi2012} and \citet{krijt2014b} with $\gamma=3/2,~Z_0=0.01,~\Sigma_\mathrm{g}(5~\mathrm{AU})=152\mathrm{~g~cm^{-2}}$, and $\alpha=10^{-3}$.

Overall, the semi-analytical model captures the growth timescales very well. The largest discrepancies are seen in the mass-range where growth is dominated by turbulent velocities, but where the grains are still relatively well coupled to the gas: $t_s < t_\eta$. In that regime (see Appendix \ref{sec:v_rel} and in particular Eq. \ref{eq:v_turb}) the relative velocity scales with the difference in Stokes numbers and is very sensitive to the choice of $\kappa$. While a closer correspondence with the MC model could be obtained by having a mass-dependent $\kappa$, we will use a constant value of $\kappa=0.5$ in the remainder of this work. The same value is used by \citet{sato2015}.

\subsubsection{Bouncing, fragmentation, and erosion}
In reality, not every collision will result in sticking and particles can bounce off each other, erode one another, or even completely fragment \citep{blumwurm2000,blumwurm2008}. The regimes in which these processes occur are usually defined by relatively sharp threshold velocities. In realistic models, the transition velocities for these collisional outcomes to occur are complex functions of the particle size(s), porosity, and material properties \citep[e.g.,][]{guttler2010}. 

Though only few experimental investigations have been performed with icy grains, it is clear that they are much stickier than their silicate counterparts \citep{dominiktielens1997,wada2013,gundlach2014}. Moreover, they are expected to grow into porous structures \citep{okuzumi2012}, making them less likely to bounce \citep{wada2011, seizinger2013a}. Therefore, we do not include bouncing and fragmentation. Recently however, \citet{krijt2014b} argued that erosion by small particles can be an effective way of halting growth when radial drift starts to play a role. We will include the effects of erosion on the growth timescale in Sect. \ref{sec:porous_erosion}.

\begin{figure*}
\centering
\includegraphics[clip=,width=.9\linewidth]{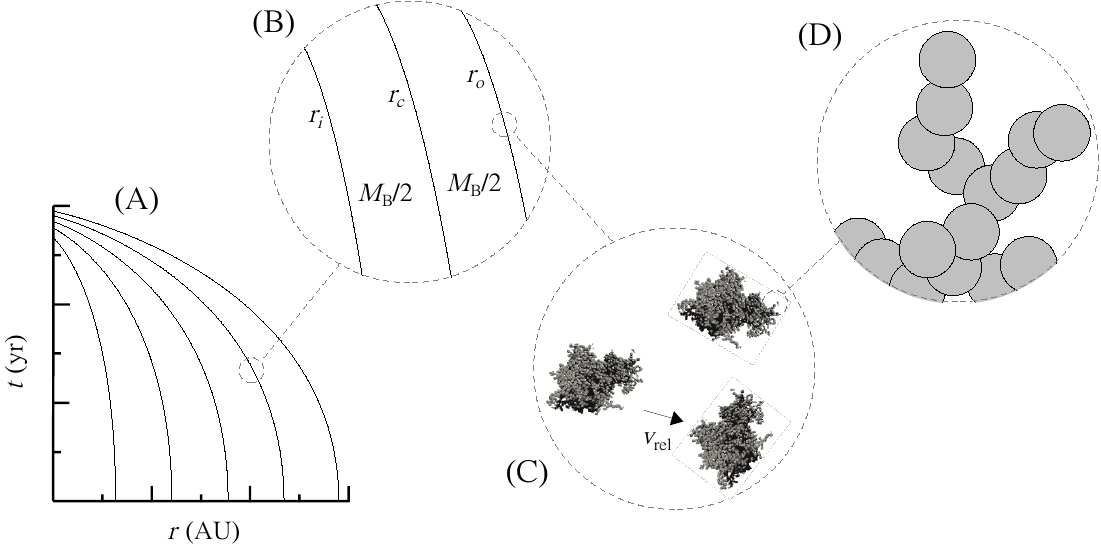}
\caption{Illustration of the batch method described in Sect. \ref{sec:batch_method}. (A): Globally, the dust content of the protoplanetary disk is described by a collection of unique and independent batches. The trajectories that individual batches follow in space and time are dubbed \emph{lifelines}. (B): Each batch is composed of three separate \emph{legs}. These legs can move closer together or further apart, but the mass enclosed between the inner two and outer two legs is always identical. For each leg, growth and radial drift are solved simultaneously using Eqs. \ref{eq:lagrange_m} and \ref{eq:lagrange_r}, while the batch's surface density and its first derivative (needed to calculate the growth timescales) are obtained using Eqs. \ref{eq:p_origin} and \ref{eq:S0}. (C): At every leg location, the local dust component is described by a monodisperse aggregate population with a single mass, size, and porosity. (D): The aggregates themselves are composed of (sub)micron-sized monomers, whose properties profoundly affect the aggregate's strength against fragmentation and compaction (see Appendix \ref{sec:phi}).}
\label{fig:schematic}
\end{figure*}

\section{The batch method}\label{sec:batch_method}
Here we introduce a new numerical approach for modeling the evolution of solids in a protoplanetary disk on a global scale. The method differs from grid-based methods \citep{brauer2007,birnstiel2010,okuzumi2012} and Monte Carlo coagulation models \citep{ormel2007,zsom2008} in a fundamental way, as it employs a Lagrangian approach, in which dust particles are followed through the disk as they grow through collisions and drift radially through the disk. A similar Lagrangian way of thinking is found in particle-tracking models that simulate diffusion in evolving disks \citep[e.g.,][]{ciesla2011,hughes2012}, but there are important differences between those methods and the approach presented here (see Sect. \ref{sec:disc_method}).

In this Section, we first introduce the principles of the method (Sect. \ref{sec:description}), describe how to include porosity and erosion (Sects. \ref{sec:include_porosity} and \ref{sec:include_erosion}), and test the method against two well-studied cases (Sect. \ref{sec:test_cases}).

\subsection{Description}\label{sec:description}
In our approach we follow the evolution of a small groups of dust particles, referred to as a \emph{batch}. The idea is to follow individual batches in time, while the constituent dust grains grow through collisions and drift inward as the result of radial drift. Specifically, we are interested in solving the Lagrangian (i.e., co-moving with the batch) derivatives for dust grains
\begin{equation}\label{eq:lagrange_m}
\frac{Dm}{Dt} = \frac{m}{t_\mathrm{grow}},
\end{equation}
and radial drift
\begin{equation}\label{eq:lagrange_r}
\frac{Dr}{Dt} = - \frac{r}{t_\mathrm{drift}},
\end{equation}
where the drift- and growth timescales are given by Eqs. \ref{eq:t_drift} and \ref{eq:t_grow}. By following a single batch in both time and space, information about the particle's history is easily accessible. The key assumption of this method is that the dust particles inside a batch grow exclusively by colliding with similar particles. Similar not only because they have comparable masses and porosities, but also because they share the same growth- and drift histories. This assumption allows us to calculate the evolution of multiple batches independently, without necessarily knowing how the dust in the rest of the disk is evolving. The validity of this assumption will be discussed further in Sect. \ref{sec:discussion}.

In order to capture the evolution of the dust surface density, we will give the batches a finite radial width. One could imagine following the front-side (slightly closer to the star) and back-side (slightly further from the star) of the batch separately. When they move closer because of small differences in the drift velocity at both sides, the surface density of dust increases; when they move away from each other, it decreases. However, this does not allow the surface density gradient to change. It transpires that we need three coordinates: one at the inner edge, one at the outer edge, and one at the center of mass. At the location of each leg, the dust is described by a mono-disperse distribution, the properties of which are initially identical, but will start to differ slightly in time because of the dependence on location in the details of the coagulation process. To create a complete picture of the dust evolution on a global scale, we will combine the results of many batches that start out at different locations. The approach is illustrated in Fig. \ref{fig:schematic}.

An alternative approach would be to calculate batches with a single leg and obtain a single global function for the surface density by looking at the distribution of all the batches simultaneously. However, in such an approach, batches cannot overtake each other, as this gives rise to infinite surface densities. An important feature of our method, is that the local surface density - determined by the spacing of a batch's three legs - is a property of the batch itself. This renders the batches truly independent, allowing one to calculate their evolution separately, but also making it possible for two batches with different constituent particles and different surface densities to occupy the same location. We will encounter such a situation in Sect. \ref{sec:porous}.

At a time $t$, the state of a batch is then fully described by the vector
\begin{equation}\label{eq:x}
\vec{x}(t) = \left( \begin{array}{ccc}
m_i \\
m_c \\
m_o \\
r_i \\
r_c \\
r_o \\
\end{array} \right),
\end{equation}
where indices $i,c,o$ correspond to the inside (or front), center of mass, and outside (or back) of the batch respectively, and by the batch's dust surface density, assumed to be a power-law with index $p$
\begin{equation}
\Sigma_\mathrm{d}(r) = \Sigma_0 \left( \frac{r}{r_c}\right)^{-p},
\end{equation}
with both $\Sigma_0$ and $p$ functions of time. At $t=0$, both $\Sigma_0$ and $p$ can be obtained from Eq. \ref{eq:Sigma_g}. Specifically, $\Sigma_0^{(t=0)} = Z_0 \Sigma_\mathrm{g}(r_c)$ and $p^{(t=0)} = \gamma$. Later, when the dust is evolving, the surface density within the batch is acquired from the locations of the three `legs' at the front, back, and center of mass of the batch. First, given that there is equal mass between $r_i$ and $r_c$ as there is between $r_c$ and $r_o$, the slope can be found by solving
\begin{equation}\label{eq:p_origin}
\int_{r_i}^{r_c} r \Sigma_\mathrm{d}(r) \intd r = \int_{r_c}^{r_o} r \Sigma_\mathrm{d}(r) \intd r,
\end{equation}
which can be approximated as
\begin{equation}\label{eq:p}
p \simeq 1 - 2 r_c  \frac{2r_c - r_i - r_o}{(r_o-r_c)^2 + (r_i-r_c)^2}.
\end{equation}
Second, the total mass inside the batch
\begin{equation}\label{eq:MB}
M_\mathrm{B} = \int_{r_i}^{r_o} 2 \pi r \Sigma_\mathrm{d}(r) \intd r = \frac{ 2\pi \Sigma_0}{p-2} \left[ r_i^{2} \left(\frac{r_i}{r_c}\right)^{-p} - r_o^{2} \left(\frac{r_o}{r_c}\right)^{-p} \right],
\end{equation}
is conserved. Since $M_\mathrm{B}$ is known at $t=0$, Eq. \ref{eq:MB} can be inverted to obtain $\Sigma_0$ at later times as a function of the location of the three legs\footnote{When $p = 2$, Equations \ref{eq:p}, \ref{eq:MB}, and \ref{eq:S0} break down. In that case, they should be replaced by $r_c = (r_i r_o)^{1/2}$ and $M_\mathrm{B}= 2 \pi r_c^2 \Sigma_0 \ln\left( r_o/r_i \right)$.}.
\begin{equation}\label{eq:S0}
\Sigma_0 = \frac{p-2}{2\pi} M_\mathrm{B} \left[ r_i^{2} \left(\frac{r_i}{r_c}\right)^{-p} - r_o^{2} \left(\frac{r_o}{r_c}\right)^{-p} \right]^{-1}.
\end{equation}
For the starting conditions, we will assume all dust particles start out as monomers, such that $m_i = m_c = m_o = m_\bullet$ at $t=0$. The initial locations of the three legs can be chosen freely, but batches should not become too wide, i.e., $(r_o-r_i)/r_c \ll 1$. 

Now everything is in place to calculate the evolution of a single batch. For a given disk model, dust properties, and batch starting conditions, the evolution of a single batch, i.e., $\vec{x}(t)$, can be calculated by integrating Eqs. \ref{eq:lagrange_m} and \ref{eq:lagrange_r} simultaneously for the three legs, while using Eqs. \ref{eq:p_origin} and \ref{eq:S0} to calculate the surface density at their locations. To solve this initial value problem, we have made use of the \texttt{scipy.integrate.ode} package of \texttt{python}. 

\begin{figure*}[!ht]
\centering
\includegraphics[clip=,width=.47\linewidth]{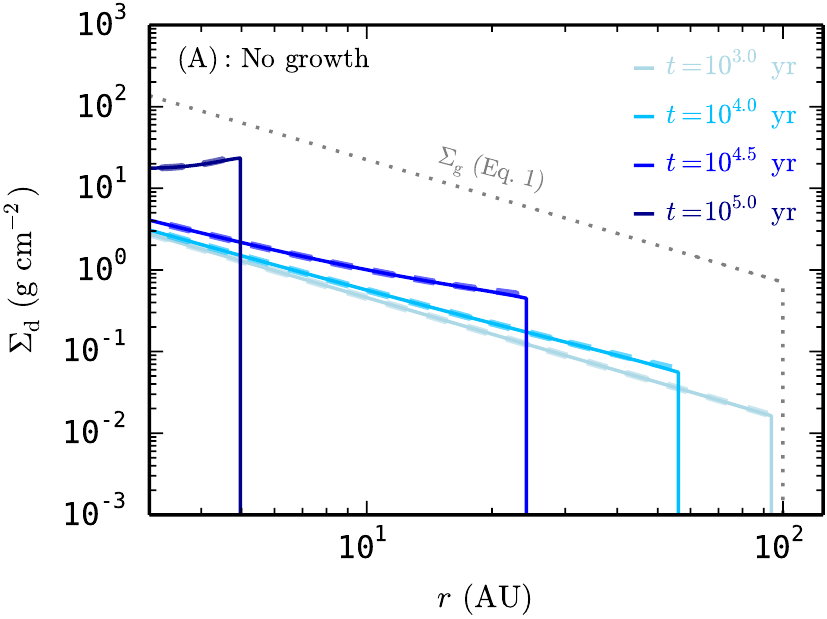}~~~~~
\includegraphics[clip=,width=.47\linewidth]{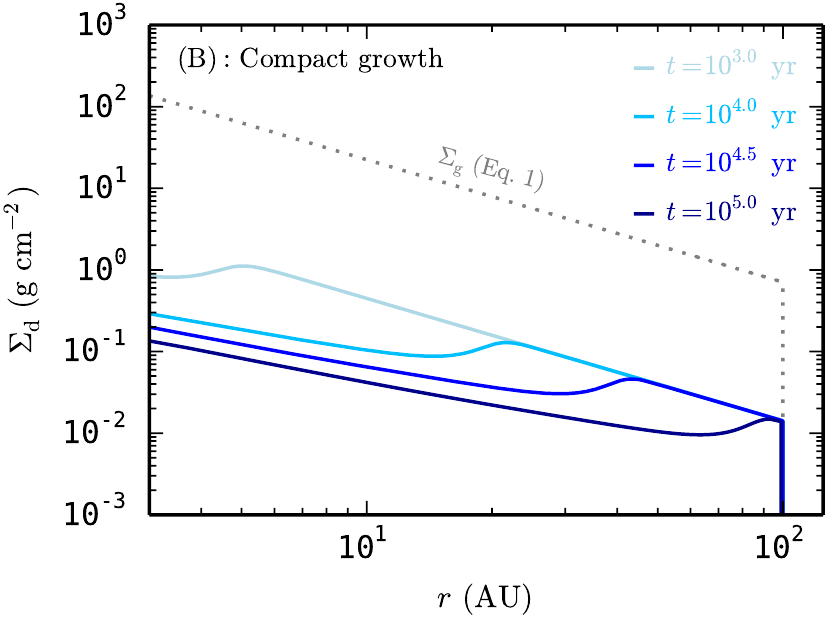}
\caption{Evolution of the dust surface density in time. (A): No growth case. Particles always have $a=1\mathrm{~mm}$. The dashed lines show analytical results of \citet{youdin2002}. (B): Compact growth with perfect sticking. Dust particles start out as $0.1\mathrm{~\mu m}$ monomers and grow on timescales given by Eq. \ref{eq:m_dot}, while staying compact at all times ($\phi=1$). The gas surface density is shown by the dotted line. The results are qualitatively different, with the no-growth case resulting in pile-ups (locally increasing $\Sigma_\mathrm{d}/\Sigma_\mathrm{g}$) and the compact coagulation model showing an inside-out removal of dust (decreasing $\Sigma_\mathrm{d}/\Sigma_\mathrm{g}$ everywhere).}
\label{fig:test_cases}
\end{figure*}

By connecting the properties\footnote{For this purpose, we use the dust properties at the batch center of mass $r_c$.} of all batches (e.g., particle mass, surface density) at specific times, the disk-wide dust distribution can be obtained as a function of time. As the batches are completely independent, there are no restrictions on their distribution. For example, it is not necessary that the edges of one batch are connected to edges of the adjacent ones. In most of this work, we will use between $10^2$ and $10^3$ batches\footnote{While the total number of batches used does not influence the evolution of the individual batches, a higher number does result in a better sampling of the disk at a given time. We found that using between $10^2$ and $10^3$ batches was sufficient to provide smooth curves in Fig. \ref{fig:test_cases}.}, distributed logarithmically between $3~\mathrm{AU}$ and $r_\mathrm{out}$.

\subsection{Treatment of porosity}\label{sec:include_porosity}
In principle, the vector $\vec{x}$ of Eq. \ref{eq:x} can be expanded to include additional dust properties such as porosity, chemical composition, water fraction, etc., provided that a recipe for determining the corresponding time-derivative exists.

In the case of porosity, obtaining an expression for $(D\phi/Dt)$ that covers fractal growth and the various different compaction mechanisms is not straightforward. Therefore, in this work, we will use a different method. Instead of adding the porosity to the vector in Eq. \ref{eq:x}, we will construct a unique function that gives the filling factor as a function of the current particle mass and location, i.e., $\phi = \phi(m,r)$. Appendix \ref{sec:phi} discusses this function in more detail. Once $\phi(m,r)$ is available, we can use it whenever we need to know a particle's porosity, for example when calculating the collisional cross section or the Stokes numbers. 

\subsection{Including erosion}\label{sec:include_erosion}
The growth rate of Eq. \ref{eq:m_dot} assumes all collisions result in sticking. However, as shown by \citet{krijt2014b}, erosive collisions have the potential to effectively halt growth when the drift velocity exceeds the threshold velocity $v_\mathrm{eros}$ needed for erosion. While the method outlined in this Section does not provide information about the population of small projectiles, we can nonetheless mimic the effect of efficient erosion by adjusting the growth timescale as follows
\begin{equation}\label{eq:t_grow_eros}
t_\mathrm{grow}^{\mathrm{(eros)}} = t_\mathrm{grow} \left(1+\exp \left\{ \left( \frac{v^*_\mathrm{rel}}{v_\mathrm{eros}} \right)^2  \right\} \right),
\end{equation}
where $v^*_\mathrm{rel}$ is the relative velocity between the particle in question and a fiducial monomer grain. In this way, the growth timescale rapidly increases when the drift velocity exceeds the erosion threshold. Both catastrophic fragmentation and bouncing could be simulated in a similar fashion, but in that case the relevant $v_\mathrm{rel}$ becomes the velocity between similar-size particles. In this work however, as we are limiting ourselves to the study of sticky ices in the outer disk, we only include erosion as a possible growth barrier.

\subsection{Test cases}\label{sec:test_cases}
In the remainder of this Section, we illustrate and test the approach outlined above by comparing it against two well-known cases: the drift-only case (neglecting coagulation completely) and the compact-growth case (without fragmentation).

\subsubsection{Test case I: drift-only}
When the particle size is kept constant in time (i.e., no coagulation takes place), grains in the outer disk drift faster than particles in the inner disks, causing pileups and potentially planetesimal formation \citep{youdin2002,youdin2004}. To study the drift-only case, we use the disk model of Table \ref{tab:benchmark} and set $\dot{m}=0$ everywhere. The dust particles are compact icy grains with a radius of $1\mathrm{~mm}$.

Figure \ref{fig:test_cases}A shows the resulting dust surface density evolution. Since, for a single grain size, Stokes numbers are largest in the outer part of the disk, particles further out will drift faster. As a result, the outer disk is slowly depleted of dust and pile-ups are created closer in \citep[e.g.,][]{youdin2002,youdin2004,birnstiel2014}. The pile-ups are caused by the slowing down of particles as they drift into the denser inner disk and are the result of the assumption of a fixed size of the particles. Below we will see that when grain sizes are limited by aerodynamical properties, no such pile-ups are created. Our model is in excellent agreement with the analytical prediction for this case \citet[][Eq. 28]{youdin2002}, shown by the overplotted dashed lines. The razor-sharp outer edge of the dust distributions in Fig. \ref{fig:test_cases}A is not actually resolved, but simply marks the location of the outermost batch at any given time. When an exponentially decaying disk is assumed instead of Eq. \ref{eq:Sigma_g}, the dust distributions at later times are smoother \citep[e.g.,][Fig. 5]{youdin2002}. The outer edge of the dust distribution will be discussed further in Sect. \ref{sec:discussion}.

\begin{figure*}
\centering
\includegraphics[clip=,width=.48\linewidth]{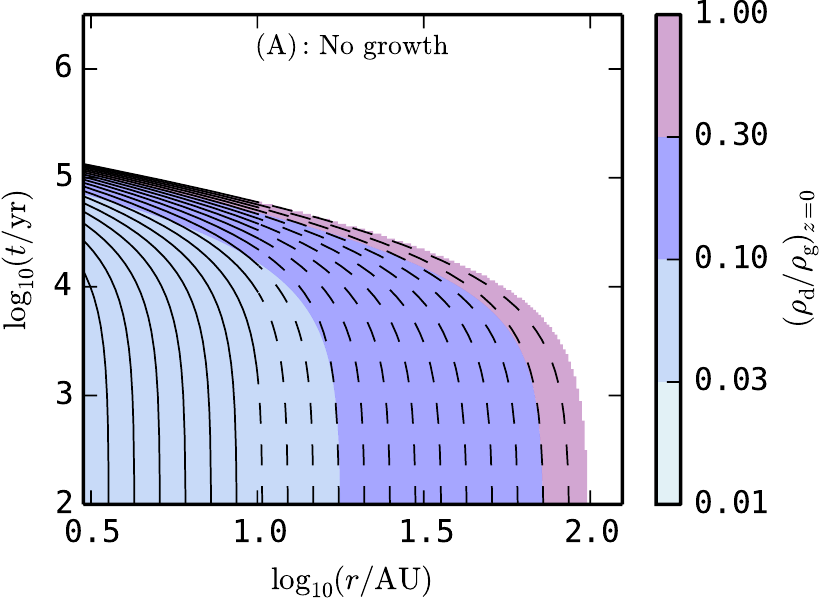}~~~
\includegraphics[clip=,width=.48\linewidth]{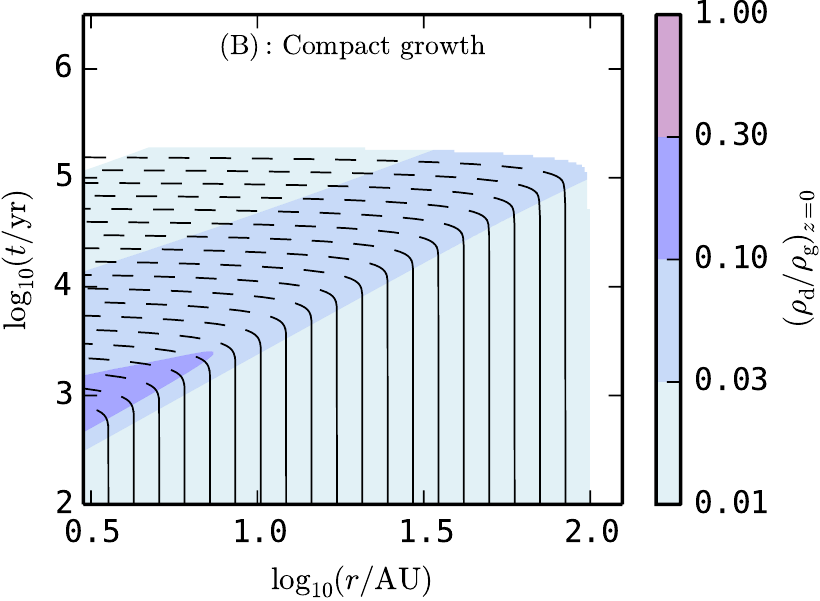}\\ \vspace{3mm}
\includegraphics[clip=,width=.48\linewidth]{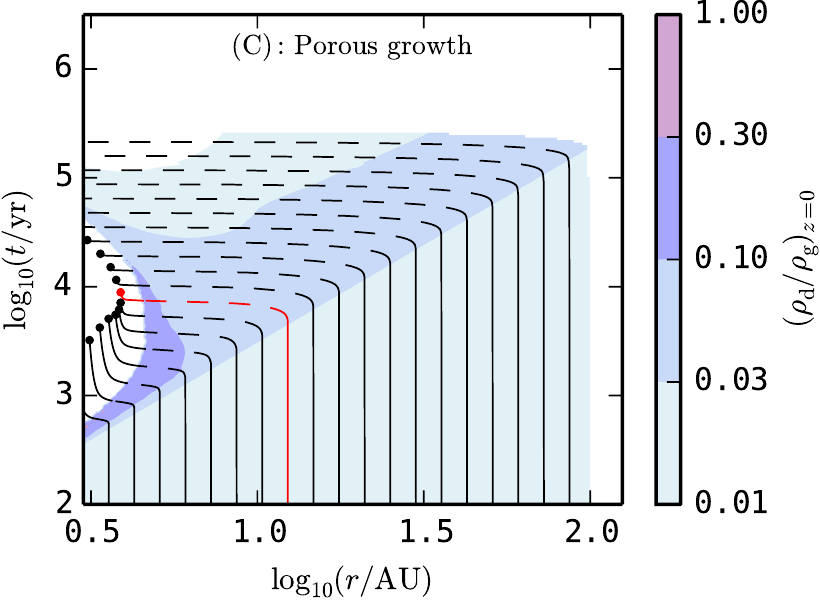}~~~
\includegraphics[clip=,width=.48\linewidth]{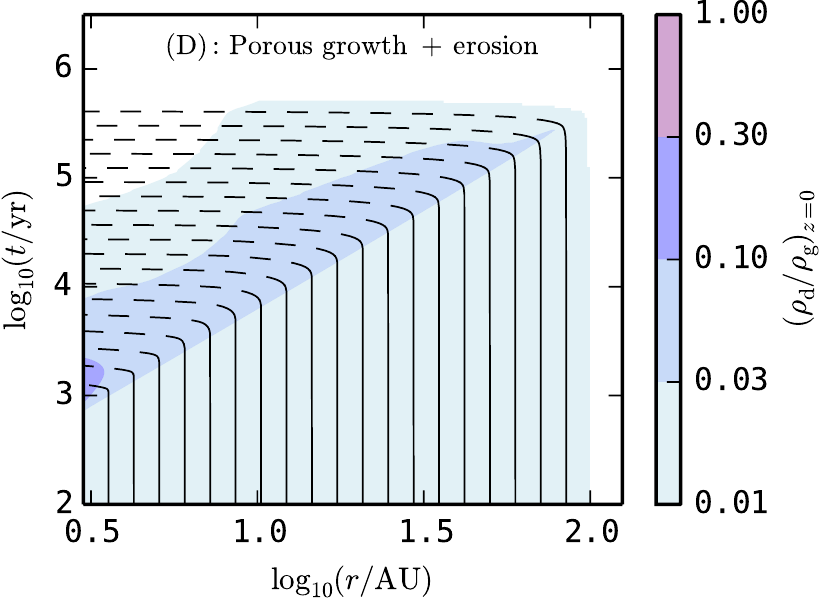}
\caption{Locations of dust batches in time for four standard cases: (A): No growth, particles are always 1 mm and compact. (B): Compact growth starting from 0.1 $\mathrm{\mu m}$ particles. (C): Porous growth with perfect sticking. Successful planetesimal formation ($\Omega t_s>10^3$) through coagulation is indicated by the ($\bullet$)-symbol. The lifeline of the batch that formed planetesimals furthest out is colored red. (D): Porous growth followed by erosion above velocities of $20\mathrm{~m~s^{-1}}$. Dashed lines indicate the particles in the batch have $10^{-2} < \Omega t_s < 3$ and background colors indicate the midplane dust-to-gas ratio of particles that have $\Omega t_s \leq 3$.}
\label{fig:panoptic}
\end{figure*}

In Fig. \ref{fig:panoptic}A, we plot the lifelines of a selection of batches for the same simulation. In this representation, batches move up in time and to the left as they start to drift in. The places where lines move close together correspond to the pile-ups in Fig. \ref{fig:test_cases}A. An important observation is that while batches get close to each other, the lifelines do not cross, justifying the assumption that they evolve independently. Dashed lines indicate the Stokes numbers of the dust particles inside the batch are $10^{-2} < \Omega t_s < 3$. Since particles are always mm-sized in the no-growth scenario, Stokes numbers are highest in the outer disk, with the transition $\Omega t_s =10^{-2}$ around $10\mathrm{~AU}$. In addition, the background colors indicate the midplane dust-to-gas ratio, calculated as
\begin{equation}\label{eq:d2g_midplane}
\left(\frac{\rho_\mathrm{d}}{\rho_\mathrm{g}}\right)_{z=0} = \frac{\Sigma_{\rm d}}{\Sigma_\mathrm{g}} \left( \frac{h_\mathrm{d}}{h_\mathrm{g}} \right)^{-1},
\end{equation}
where the fraction of scale-heights on the right-hand-side is a function of the particle Stokes number (see Eq. \ref{eq:h_d}). Initially, because of settling, the highest midplane dust-to-gas ratios are found exclusively in the outer disk. Later, when pile-ups are formed,  high dust-to-gas ratios are reached at smaller radii as well. After ${\sim}10^5\mathrm{~yr}$, all batches have drifted inside the snow-line and the solid content of the disk has vanished. 

\subsubsection{Test case II: compact growth}\label{sec:test_case_II}
The next test includes growth, but assumes that the particles stay compact ($\phi=1)$ at all times. For this test, the dust particles start out as monomers with $a_\bullet=0.1\mathrm{~\mu m}$ and their growth timescale is calculated using Eq. \ref{eq:m_dot}. Disk-wide numerical simulations of compact coagulation (with or without fragmentation) and drift have been performed by many authors, including \citet{brauer2007,birnstiel2010, okuzumi2012, birnstiel2014}.

Figures \ref{fig:test_cases}B and \ref{fig:panoptic}B show the results of the panoptic model for compact growth in the disk of Table \ref{tab:benchmark}. The behavior of the dust component is fundamentally different from the no-growth case in a number of ways. We focus on Fig. \ref{fig:panoptic}B first. Since particles start out as well-coupled sub-micron monomers, $\Omega t_s \ll 10^{-2}$ everywhere in the disk and radial drift is negligible. However, Brownian motion and the presence of turbulence causes the particles to collide and grow. As the grains gain mass, they slowly decouple from the gas and their Stokes number increases. At some point, the drift timescale becomes comparable to the growth timescale and the particles will start to drift inward. Because the timescales for growth are shortest at small radii, grains that start out in the inner disk reach high Stokes numbers much earlier. This causes an \emph{inside-out} clearing of the dust component, very different from the no-growth scenario. As illustrated in Fig. \ref{fig:test_cases}B, the inside-out removal of dust does not allow the formation of pile-ups and the dust surface density decreases in time at every location.

For turbulence-driven growth in the Epstein drag regime (as is the case here), a pure drift-growth balance leads to a dust surface density profile $\Sigma_\mathrm{d} \propto ( \Sigma_\mathrm{g} r^{-2} \Omega^{-2})^{1/2}$ \citep[e.g.,][]{birnstiel2012}. For $\gamma=3/2$, this results in $\Sigma_\mathrm{d} \propto r^{-1}$. The surface density curves of Fig. \ref{fig:test_cases}B are in excellent agreement with this prediction and with grid-based numerical models of similar disks \citep[e.g.,][Fig. 3]{okuzumi2012}.

\section{Porous growth with perfect sticking}\label{sec:porous}
While compact growth typically results in an inside-out removal of the disk's solids, highly porous growth could potentially lead to planetesimal formation by allowing fluffy icy grains to quickly grow through the radial drift barrier \citep{okuzumi2012}. In this Section, we will study such highly porous growth using our Lagrangian approach and quantify in which regions of the protoplanetary disk such rapid growth is feasible.

The set-up is the same as in Sect. \ref{sec:test_case_II}, but the dust particle porosity $\phi(m,r)$ is now calculated assuming hit-and-stick growth followed by compaction by gas-pressure and eventually self-gravity (Sect. \ref{sec:include_porosity} and Appendix \ref{sec:phi}). When the aggregates inside a batch grow to Stokes numbers $\Omega t_s > 10^3$, we classify them as planetesimals\footnote{At $5\mathrm{~AU}$, these Stokes numbers correspond to aggregates with a mass of about ${\sim}10^{14}\mathrm{~g}$ and a porosity of ${\sim}1\%$.} and no longer follow their evolution.

Figure \ref{fig:panoptic}C shows the result of the porous growth case assuming $0.1\mathrm{~\mu m}$ icy monomers and perfect sticking. The ($\bullet$)-symbols indicate successful planetesimal formation. Our results confirm the findings of \citet{okuzumi2012}, \citet{kataoka2013c} and \citet{krijt2014b}, indicating that growth through the radial drift barrier is possible in the region just outside the snow-line. Indeed, dust can grow from sub-micron to planetesimal sizes within ${\sim}10^4\mathrm{~yr}$. In the outer disk, the behavior is similar to that in Fig. \ref{fig:panoptic}(B): dust grains grow to $\Omega t_s \sim 1$ and drift inward. We note however, that in the porous case these $\Omega t_s =1$ grains have masses and porosities that differ from the compact grains by many orders of magnitude \citep[see Fig. 3 of][]{johansen2014}.

\subsection{Planetesimals and pebbles}\label{sec:planetesimals_pebbles}
Using Fig. \ref{fig:panoptic}C, three distinct regions can be identified inside the protoplanetary disk. First, we identify the batch that formed the outermost planetesimals. The lifeline of this batch is shown in red in Fig. \ref{fig:panoptic}C. Dust particles that are located inside of this batch at $t=0$ will eventually grow into planetesimals on a timescale of ${\sim}10^4\mathrm{~yr}$.

Material located outside of the red batch at $t=0$ will grow to $\Omega t_s \sim 1$ and drift in. We call these particles pebbles\footnote{While some batches starting outside of the red batch form planetesimals, these grains have at some point drifted past the already-formed planetesimals in the inner disk.}. Around $10^4\mathrm{~yr}$, material from the outer disk will start to drift into the planetesimal belt located just behind the snow-line. Since our model at this point does not include any interaction between batches, the pebbles fly through the planetesimals unaffected. In reality, the planetesimals can potentially accrete the pebbles very efficiently \citep{ormelklahr2010,guillot2014}. In Fig. \ref{fig:panoptic}C, material that starts out outside ${\sim}10\mathrm{~AU}$ will form pebbles. For the disk profile of Eq. \ref{eq:Sigma_g}, the fraction of the disk's total mass located outside $r$ equals
\begin{equation}
\mathcal{R} \equiv \dfrac{ \int_{r}^{r_\mathrm{out}} r \Sigma_\mathrm{d}(r) \intd r }{ \int_{0}^{r_\mathrm{out}} r \Sigma_\mathrm{d}(r) \intd r } = 1-\left( \frac{r}{r_\mathrm{out}} \right)^{2-\gamma}.
\end{equation}
Thus, the pebbles formed outside of $10\mathrm{~AU}$ together hold ${\sim}2/3$ of the total dust mass and have the potential to boost protoplanet growth after the formation of planetesimals. We will discuss pebble accretion further in Sect. \ref{sec:discussion}.

\subsection{Influence of disk mass}
\begin{figure}
\centering
\includegraphics[clip=,width=.95\linewidth]{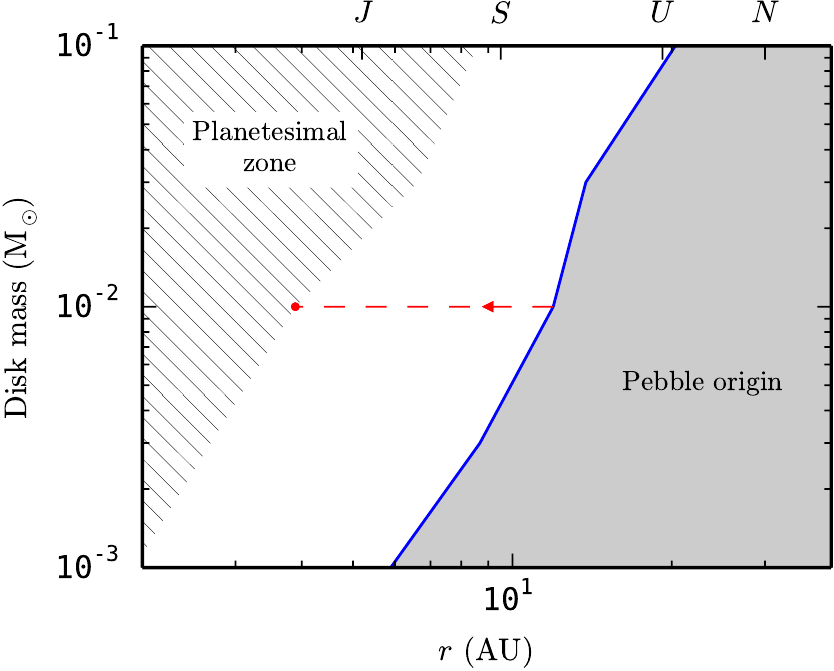}
\caption{The influence of total disk mass on the location where planetesimals and pebbles form (Sect. \ref{sec:planetesimals_pebbles}). Material that starts out left of the solid blue line successfully grows through the drift barrier, eventually forming planetesimals a bit closer in. The red outermost planetesimal of Fig. \ref{fig:panoptic}C has been plotted for the $M_\mathrm{D}/M_\odot=10^{-2}$. Material that starts out in the gray region cannot coagulate fast enough, experience rapid radial drift, and at some point drift past the already-formed planetesimals in the inner disk. For reference, the current locations of the four Jovian planets are indicated by capital letters.}
\label{fig:planzone}
\end{figure}

In Fig. \ref{fig:planzone} we combine simulations performed with different disk masses (the other properties, i.e., turbulence, disk outer radius, metallicity, are identical to those in Table \ref{tab:benchmark}) and show how disk mass influences the location where planetesimals and pebbles originate. The figure should be read as follows: For a given disk mass, material that starts to the left of the solid blue line will form planetesimals in the `planetesimal zone' marked by the shaded region. Material that starts out in the gray area will grow and drift to enter the previously formed planetesimal zone at Stokes numbers $10^{-2} < \Omega t_s < 3$. The red lifeline of Fig. \ref{fig:panoptic}C is also plotted at the corresponding disk mass of $10^{-2}M_\odot$. 

From the figure we can identify two trends. First, more massive disks form planetesimal at larger radii. This is to be expected, as more massive disks have a higher dust surface density, making growth through the radial drift barrier easier \citep[e.g.,][Sect. 4]{okuzumi2012}. For disk masses comparable to the MMSN, planetesimal formation around the current location of Jupiter is a realistic possibility. Second, the boundary between the regions where planetesimals and pebbles originate, moves out for increasing disk mass. This means that for more massive disks, a larger fraction of the dust content can form planetesimals directly, leaving a smaller fraction available to be accreted in the form of pebbles later. Nonetheless, for the most massive disks considered here the pebble origin region still contains ${\sim}1/2$ of all the dust.

\section{Porous growth with erosion}\label{sec:porous_erosion}
So far we have assumed that all collisions result in perfect sticking. And while collisions between icy same-size aggregates in the outer disk are indeed likely to occur below the critical fragmentation threshold, \citet{krijt2014b} have shown that high-velocity collisions with large mass ratios can result in efficient erosion and possibly frustrate the further growth of aggregates with Stokes numbers $\Omega t_s\sim1$. Here, we mimic the effect of efficient erosion by adjusting the growth timescale according to Eq. \ref{eq:t_grow_eros}, using an erosion threshold velocity of $20\mathrm{~m~s^{-1}}$ \citep{gundlach2014}.

The results of the simulation with erosion are shown in Fig. \ref{fig:panoptic}D. The largest difference with the perfect sticking scenario of Fig. \ref{fig:panoptic}C is that there are no planetesimals being formed in the inner disk. Instead, when particles grow to $\Omega t_s \sim 1$, their growth is frustrated by erosion and they drift inward.

\subsection{Streaming instability}\label{sec:SI}
Apart from coagulating directly, planetesimals can also be formed through streaming instability (SI) \citep[e.g.,][]{youdin2005,johansen2007N,bai2010a,bai2010b}. In this section we investigate how porous coagulation followed by erosion can lead to conditions suitable for triggering SI.

\begin{figure*}
\centering
\includegraphics[clip=,height=.27\linewidth]{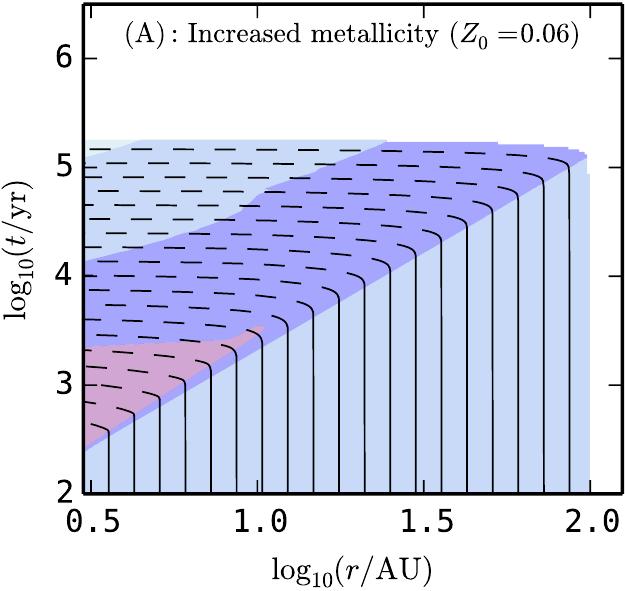}~~
\includegraphics[clip=,height=.27\linewidth]{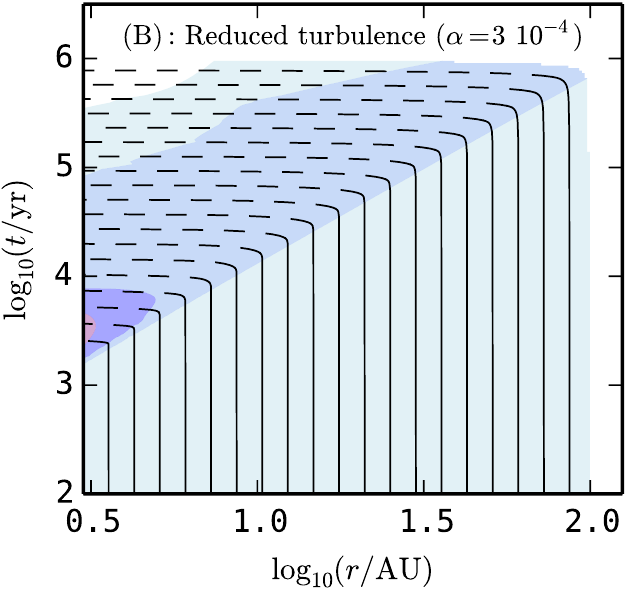}~~
\includegraphics[clip=,height=.2745\linewidth]{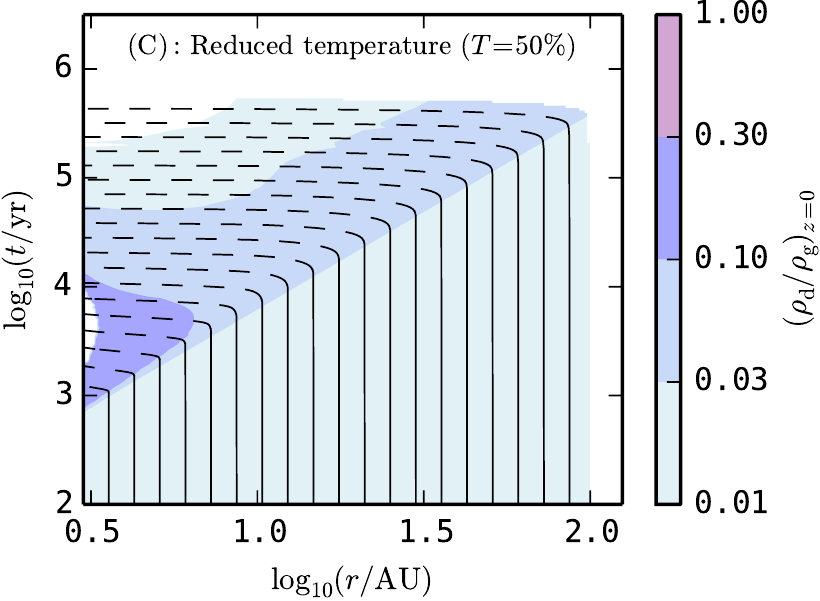}
\caption{Similar to Fig. \ref{fig:panoptic}D (porous growth + erosion), but with (A) an increased metallicity of $Z_0=0.06$; (B) a reduced turbulence strength of $\alpha=3\times10^{-4}$; and (C) a gas temperature 50\% lower than that in Eq. \ref{eq:T}. All three modifications result in a higher maximum dust-to-gas ratio compared to Fig. \ref{fig:panoptic}D.}
\label{fig:SI}
\end{figure*}

We make use of the work of \citet{drazkowska2014b}, who, based on the work of \citet{johansen2007N,johansen2009} and \citet{bai2010a,bai2010b} defined three conditions for triggering SI. For a mono-disperse particle distribution, these conditions are equivalent to:
\begin{enumerate}
\item[$i)$]{The Stokes number of the mass dominating particles needs to be close to unity. Typically, one needs $10^{-2} \leq \Omega t_s \leq 3$.}
\item[$ii)$]{The midplane dust-to-gas ratio of these particles needs to exceed, or be close to, unity.}
\item[$iii)$]{The vertically integrated metallicity should be at least a few times Solar, i.e., $Z \gtrsim 0.02-0.03$.}
\end{enumerate}
See also Fig. 8 of  \citet{carrera2015} for points $i)$ and $iii)$. Lastly, SI needs time to develop and will not be triggered if the growth timescales of the particles are too short. In other words, $t_\mathrm{grow} > \Omega^{-1}$. The first two conditions are related to efficient momentum transfer between dust particles and the gas. Particles with much higher Stokes numbers do not effectively interact with the gas, while particles with with much smaller stopping times do not result in strong clumping. Condition \emph{iii)} is related to suppressing midplane turbulence. For low metallicities, the strength of this turbulence drops sharply \citep{bai2010a}.

Concentrating on conditions \emph{i)} and \emph{ii)} and armed with the semi-analytical model of Sect. \ref{sec:batch_method}, we can identify regions in space and time where these conditions are met. In Fig. \ref{fig:panoptic}, dashed lines indicate the particles in the batch meet condition \emph{i)}, while the background colors show the midplane dust-to-gas ratio (Eq. \ref{eq:d2g_midplane}), important for condition \emph{ii)}.

Focussing on the porous growth + erosion case (Fig. \ref{fig:panoptic}D), we see that conditions for SI are not reached. While every batch eventually forms particles with high Stokes numbers (evidenced by the dashed lines), the achieved midplane densities are about a factor 10 too low. We can see a \emph{front} of moderate midplane densities moving outward in time, starting in the inner disk at $10^3\mathrm{~yr}$ and reaching $100\mathrm{~AU}$ after almost $10^6\mathrm{~yr}$. 

This behavior can be understood in the following way. Essentially, the midplane dust-to-gas ratio is the combination of the vertically integrated metallicity and the degree of settling
\begin{equation}\label{eq:d2g_approx}
\left(\frac{\rho_\mathrm{d}}{\rho_\mathrm{g}}\right)_{z=0} \sim Z  \left( \frac{\Omega t_s}{\alpha}\right)^{1/2}.
\end{equation}
where $Z=(\Sigma_\mathrm{d} / \Sigma_\mathrm{g})$ and we have used that $h_\mathrm{d}/h_\mathrm{g}\sim(\alpha/\Omega t_s)^{1/2}$ for high Stokes numbers (see Eq. \ref{eq:h_d}). Second, since the removal of dust occurs from the inside-out, the surface density -- at a given $r$ -- will decrease, very much like in Fig. \ref{fig:test_cases}B, and pile-ups like the ones seen in the no-growth case do not occur. As a result, the highest vertically integrated metallicity is achieved at the very beginning and $Z \leq Z_0$. The highest midplane densities are then achieved in the interval when growth toward $\Omega t_s > 10^{-2}$ has occurred, but before enough time has past for radial drift to reduce the local dust surface density. The diagonal shape of high $(\rho_\mathrm{d}/\rho_\mathrm{g})_{z=0}$ in Fig. \ref{fig:panoptic}D occurs because growth takes longer in the outer disk.

\subsection{Disk metallicity, turbulence, and temprature}\label{sec:SI_help}
Inspired by Eq. \ref{eq:d2g_approx}, we briefly discuss three ways that can help to create conditions suitable for SI. All three are shown in Fig. \ref{fig:SI}.

\emph{Increasing the initial dust content:} The most straightforward way to reach higher midplane dust-to-gas ratios is to start out with a higher metallicity from the beginning. Fig. \ref{fig:SI}A shows that for $Z_0=0.06$, a mass loading close to unity can be achieved in the inner ${\sim}10\mathrm{~AU}$.

\emph{Decreasing the turbulence strength:} The turbulence strength influences the degree of settling and for a given metallicity and particle size, the midplane dust-to-gas ratio is inversely proportional to $\sqrt{\alpha}$. For example, reaching a mass loading of 1 in a column with $Z=0.01$ and $\Omega t_s=0.1$ particles requires $\alpha \sim 10^{-5}$. However, the turbulence cannot be decreased to arbitrarily small values. The presence of marginally decoupled grains will give rise to Kelvin-Helmholtz instability \citep{weidenschilling1980,weidenschilling1995,johansen2006} and SI \citep[e.g.,][]{takeuchi2012}, resulting in turbulence that can be parametrized with for example Eq. 8 of \citet{drazkowska2014b}. For the disk models used in this work, the strength of this turbulence amounts to $\alpha_\mathrm{mp}\simeq 5\times10^{-4}$ and ${\sim}10^{-3}$ at 5 and $50\mathrm{~AU}$, respectively, for particles with Stokes numbers around unity.

\emph{Decreasing the temperature:} The temperature structure used so far (Eq. \ref{eq:T}) is based on an optically thin disk. However, midplane temperatures in disks might be significantly lower, especially if the disk is optically thick \citep[e.g.,][]{andrews2009}. In contrast, the presence of viscous heating can affect the temperature profile significantly \citep{bitsch2015}. Lowering the temperature affects many things. Most important for the calculations presented here is that a lower temperature decreases the pressure gradient $\eta$ (see Eq. \ref{eq:eta}), causing the radial drift to be slower. This makes erosion less efficient (lower $\eta v_\mathrm{K}$) and also gives grains more time to grow, allowing them to reach larger sizes. Reducing the temperature by 50\% indeed results in higher midplane dust-to-gas ratios (Fig. \ref{fig:SI}C), but the effect is not large enough to trigger SI.

\section{Discussion}\label{sec:discussion}

\subsection{Batch approach}\label{sec:disc_method}
The numerical approach introduced in this work (Sect. \ref{sec:batch_method}), in which individual batches of dust are followed as they grow and drift in the protoplanetary nebula, provides an intuitive and flexible way to calculate (porous) coagulation and the evolution of the global surface density simultaneously, connecting the properties of the (sub)micron monomers to the growing aggregates and the global redistribution of solids (Fig. \ref{fig:schematic}). The flexibility and speed of this approach allow us to study the impact of different coagulation models (Fig. \ref{fig:panoptic}) and or disk properties (Fig. \ref{fig:SI}) quickly, while preserving the essential characteristics of the growth process. 

At this point, the method has three main drawbacks. First, it traces only the mass-dominating particles and does not provide information about the number distribution for smaller masses. If the distribution can be assumed to be in growth/fragmentation equilibrium, the complete mass distribution may be reconstructed \citep[e.g.,][]{birnstiel2011}, though this has not yet been attempted for porous growth or for a steady state between growth and erosion. 

Second, batches do not interact with each other. In most situations encountered here this is justified, as evidenced by the fact that the different lifelines do not cross in Fig. \ref{fig:panoptic}. In some cases however, this assumption does not hold. One example, in which pebbles from the outer disk drift past planetesimals in the inner disk, is found in Sect. \ref{sec:porous} and will be discussed below. Another situation where batches come uncomfortably close is in the outer edge of an exponential disk. \citet{birnstiel2014} have shown that in exponentially decaying disks, a pile-up will be created at the outer edge (e.g., their Figs. 3 and 4). In theory, these pile-ups host \emph{all} dust grains that were originally located further out. Coagulation inside such a pile-up would be hard to model with independent dust batches. 

Third, the approach does not include radial mixing as a result of turbulent diffusion. With one of the underlying assumptions of the model being that dust grains that start out together evolve in a similar manner, it appears not to be straightforward to add these processes. Radial mixing and diffusion are accurately captured by particle tracking models that simulate individual trajectories for a large number of individual grains \citep[e.g.,][]{ciesla2011,hughes2012}. Such models differ from the method presented here in a number of ways. First, in our approach, the dust surface density is solved from the evolution of a batch itself, rather than being obtained by looking at the dynamical evolution of all the particles in the simulation. Second, where we focus on the effects of the details of the coagulation process in a static disk, these works primarily look at evolving gas disks and either neglect particle coagulation \citep{ciesla2011} or treat it in a simplified way; for example by injecting particles of increasing size at a later stages \citep{hughes2012}. Despite these differences, the main conclusions of \citet{hughes2012} and this work are similar: both do not find significant enhancements of the vertically-integrated dust-to-gas ratio.




\subsection{Rapid porous growth}
For porous icy aggregates, fragmentation occurs only at very high velocities \citep{wada2013} and the resulting fluffy aggregates can swiftly grow through the radial drift barrier (Fig. \ref{fig:panoptic}C). Rapid growth is harder to achieve in the outer disk, since timescales are longer and dust and gas spatial densities are lower. As a result, planetesimals form in a zone just behind the snow-line. Figure \ref{fig:planzone} shows how the size of this region depends on disk mass. Comparing the planetesimal formation zone to the current locations of the Solar System giant planets\footnote{In the \emph{Nice} model however, the giant planets of the Solar System were originally located much closer in, roughly between 5 and 17 AU, migrating out at later times \citep{tsiganis2005, morbidelli2005, gomes2005}.}, we find that while it is hard to form planetesimals all the way out to the current location Uranus and Neptune, it is very well possible to form planetesimals in the location where Jupiter is now. One could envision a scenario where enough planetesimals accumulate to trigger the formation of Jupiter early on. At that point, the assumptions in our model break down and the presence of Jupiter will steer the evolution of the disk and planet formation therein \citep[e.g.,][]{pollack1996,pinilla2011,kobayashi2012}. 

\subsection{Pebble accretion}
After being studied by \citet{ormelklahr2010,lambrechts2012}, pebble accretion has received a lot of attention as a robust way of growing planetesimals and planetary embryos \citep{lambrechts2014, kretke2014, levison2015}. The method developed in this work is ideally suited for studying pebble accretion. First, it provides the location and formation time of the planetesimal population (e.g., Fig. \ref{fig:panoptic}C), but it also accurately captures the properties of the pebbles that drift in later and provides information about their history (e.g., where the pebbles originated from, how long they spent in which part of the disk, etc.). The next step will be to include interaction between different batches when they overlap. Once could imagine syphoning mass from the batch with pebbles to the batch with planetesimals with a certain efficiency. This accretion efficiency depends sensitively on the properties of the local gas, the size of the planetesimals, and the aerodynamic properties of the pebbles \citep[e.g.,][]{guillot2014,visser2015}, information that is all readily available in our approach.

\subsection{Erosion and streaming instability}
In \citet{krijt2014b}, it was seen that the growth of $\Omega t_s \sim 1$ particles stagnates (i.e., $\dot{m} \rightarrow 0$) soon after the drift velocity exceeds the erosion threshold velocity. The growth stagnated because a balance was reached between growth through similar-size collisions and erosion by small grains \citep[][Fig. 10]{krijt2014b}. Inspired by these results, we adopted Eq. \ref{eq:t_grow_eros} to simulate efficient erosion. While instructive, this treatment of erosion is very rudimentary, but since our monodisperse approach does not hold information about the population of small grains, a self-consistent treatment of erosion is not possible at this point. Unfortunately at the moment, no global codes are available that can self-consistently treat the porosity evolution and radial drift of the full mass distribution while taking into account destructive and erosive collisions.

In Sect. \ref{sec:SI}, we investigated when porous growth might lead to SI, making use of the conditions as defined by \citet{drazkowska2014b}. Our simulations indicate that, in a smooth disk, reaching conditions for SI is not straightforward. While porous growth (possibly followed by erosion) naturally leads to a population of aggregates with $10^{-2} \leq \Omega t_s \leq 3$ in a large portion of the protoplanetary disk (Figs. \ref{fig:panoptic}C and \ref{fig:panoptic}D), the creation of a dense midplane layer of solids is problematic. The main reason for this is that growth followed by drift inevitably results in an inside-out removal of the solid content and does not lead to pile-ups (see also Fig. \ref{fig:test_cases}). These results are in line with \citet{drazkowska2014b}, who, focussing on compact growth, concluded that SI can only follow if the bouncing barrier can be overcome, the local metallicity can somehow be enhanced, and the turbulence is sufficiently weak. A number of ways to increase the maximum midplane dust-to-gas ratio are discussed in Sect. \ref{sec:SI_help} and depicted in Fig. \ref{fig:SI}. Such variations in the disk properties, perhaps working together, could result in conditions suitable for SI in a subset of protoplanetary disks. 

\subsection{Sintering}
One intriguing possibility of creating regions of enhanced dust density is by having dust sintering in specific regions of the disk. Sintering is expected to lower the fragmentation threshold velocity in regions around the ice-lines of major volatile species \citep{sirono2011}, effectively creating alternating regions of fragmentation-limited and drift-limited growth. Recently, \citet{okuzumi2015} have argued that the existence of such sintering regions can result in pile-ups of material, offering a possible explanation for the peculiar shape of the HL Tau protoplanetary disk \citep{alma2015}. While it appears sintering can play an important role in the radial redistribution of solids in some systems, there are still many uncertainties. For example, it is not yet clear how sintering affects the porosity evolution described in Appendix \ref{sec:phi} or how exactly it alters the expected collisional outcomes.

\section{Conclusions}\label{sec:conclusions}
We have developed a novel Lagrangian approach for calculating the evolution of the mass-dominating dust aggregates as they grow and drift in a protoplanetary disk. The method, summarized in Fig. \ref{fig:schematic}, allows the calculation of the global evolution of the dust surface density on Myr timescales while preserving the essential characteristics of the porous growth process and can be used to study planetesimal formation and pebble delivery.

After testing the new approach against two well-known cases (Fig. \ref{fig:test_cases}), we use it to study the formation of the first generation of planetesimals -- those that can form in a smooth disk structure -- in disks around Sun-like stars. When fragmentation and erosion are inefficient, we find that:
\begin{enumerate}
\item{Planetesimals can coagulate very rapidly, within ${\sim}10^4\mathrm{~yr}$, around the current location of Jupiter. While they end up in the region just behind the snow-line, the planetesimals include material from a broader region that extends out to ${\sim}10\mathrm{~AU}$ for an MMSN-like disk (Fig. \ref{fig:panoptic}C).}
\item{For more massive disks, both the region where planetesimals eventually form and the region where they originate from move outward (Fig. \ref{fig:planzone}). Thus, in these massive disks, a larger fraction of the dust content can directly form planetesimals, leaving less material to be accreted as pebbles later on.}
\end{enumerate}

\noindent These scenarios rely on the assumption that fragmentation and erosion are relatively unimportant at collision speeds up to several tens of $\mathrm{m~s^{-1}}$. Alternatively, when erosion balances growth around $\Omega t_s \sim 1$ \citep{krijt2014b}, further coagulation is not possible but conditions necessary for streaming instability (SI) might be reached. In these cases, we find that:
\begin{enumerate}
\item[3.]{While porous growth limited by drift-induced erosion is an effective way of creating aggregates with $10^{-2} \leq \Omega t_s \leq 3$ in a large region of the protoplanetary disk (Figs. \ref{fig:panoptic}D and \ref{fig:SI}), conditions needed for SI are generally not reached.}
\item[4.]{The most stringent condition is creating and maintaining a dense midplane layer of solids. In a smooth gas disk, rapid porous growth followed by erosion leads to an inside-out clearing of the dust disk. In such a scenario, no pile-ups are created and the metallicity decreases.}
\item[5.]{The highest midplane densities are reached in the inner disk first and then move out toward the outer parts. A reduced turbulence level and lower gas temperature increase the maximum midplane dust-to-gas ratios slightly, but the biggest impact comes from increasing the initial metallicity (Fig. \ref{fig:SI}).}
\end{enumerate}

\noindent Future improvements to the method, in particular the addition of interaction between batches, will help build a coherent picture of the planet formation process.

\begin{acknowledgements}
Dust studies at Leiden Observatory are supported through the Spinoza Premie of the Dutch science agency, NWO. The authors would like to thank C.\,P.~Dullemond, A.~Johansen, J.~Dr{\c a}{\.z}kowska, T.~Birnstiel and S.~Okuzumi for encouraging discussions and the anonymous reviewer for comments that helped improve the manuscript.
\end{acknowledgements}

\bibliographystyle{aa}
\bibliography{refs}

\begin{appendix}

\section{Particle stopping time}\label{sec:t_s}
The particle stopping time is a function of an aggregate's mass $m$ and size $a$, and the properties of the local gas. Depending on the aggregate size in relation to a gas molecule mean free path $\lambda_\mathrm{mfp}$, the stopping time is given by the Epstein or Stokes drag regime through
\begin{equation}\label{eq:t_s}
t_s = \begin{cases} 
~t_s^{\mathrm{(Ep)}} = \dfrac{3m}{4\rho_\mathrm{g} v_{\mathrm{th}}A}  &\textrm{~~for~~} a < \dfrac{9}{4}\lambda_{\mathrm{mfp}},\vspace{3mm} \\
~t_s^{\mathrm{(St)}} = \dfrac{4a }{9 \lambda_{\mathrm{mfp}}}t_s^{\mathrm{(Ep)}}  &\textrm{~~for~~} a > \dfrac{9}{4}\lambda_{\mathrm{mfp}},
\end{cases}
\end{equation}
with $v_{\mathrm{th}} = \sqrt{8/\pi}c_\mathrm{s}$ the mean thermal velocity of the gas molecules. The mean free path depends on the gas density and is given by  $\lambda_{\mathrm{mfp}} = m_\mathrm{g}/(\sigma_{\mathrm{mol}}\rho_\mathrm{g})$, with $\sigma_{\mathrm{mol}}=2\times10^{-15}\mathrm{~cm^{2}}$ the molecular cross section. For porous aggregates, the cross section $A$ equals the orientation-averaged projected cross-section \citet{okuzumi2009}.

Eq. \ref{eq:t_s} is valid when the particle Reynolds number $\mathrm{Re_p}= 4 a v_\mathrm{dg}/(v_\mathrm{th} \lambda_\mathrm{mfp}) <1$, with $v_\mathrm{dg}$ the relative velocity between the gas and the dust particle. For the largest bodies however, this condition is often not met. In these cases, it is useful to write the stopping time as
\begin{equation}
t_s = \frac{2m}{C_\mathrm{D} \rho_\mathrm{g} v_\mathrm{dg} A},
\end{equation}
with $C_\mathrm{D}$ the drag coefficient. Following \citet{weidenschilling1977}, we use
\begin{equation}\label{eq:newton}
C_\mathrm{D} = \begin{cases}
~24(\mathrm{Re_p})^{-1} &\textrm{~for~~} \mathrm{Re_p}<1,\vspace{3mm} \\
~24(\mathrm{Re_p})^{-3/5} &\textrm{~for~~} 1<\mathrm{Re_p}<800,\vspace{3mm} \\
~0.44&\textrm{~for~~} 800<\mathrm{Re_p}.
\end{cases}
\end{equation}
For large Reynolds numbers, the stopping time depends the velocity relative to the gas, and one has to iterate to obtain $t_s$.

\section{Particle relative velocity}\label{sec:v_rel}
The relative velocity between particle 1 and particle 2 is obtained by adding various velocity sources quadratically. We take into account relative velocities arising from Brownian motion, turbulence, radial drift, and azimuthal drift. The Brownian motion relative velocity is given by
\begin{equation}
\Delta v_\mathrm{BM} = \sqrt{\frac{8 k_\mathrm{B} T (m_1 + m_2)}{\pi m_1 m_2}},
\end{equation}
and depends only on particle masses and temperature.

The turbulence-induced relative velocity between two particles with stopping times $t_{s,1}$ and $t_{s,2}\leq t_{s,1}$ has three regimes \citep{ormel2007b}
\begin{equation}\label{eq:v_turb} 
\Delta v_\mathrm{turb} \simeq \delta v_\mathrm{g} \times \begin{cases}
~\mathrm{Re_t}^{1/4} \ \Omega (t_{s,1}-t_{s,2}) &\textrm{~for~} t_{s,1} \ll t_\eta,\vspace{3mm} \\
~1.4\dots1.7 \left(\Omega t_{s,1} \right)^{1/2} &\textrm{~for~} t_\eta \ll t_{s,1} \ll \Omega^{-1}, \vspace{3mm} \\ 
~\left(\dfrac{1}{1+\Omega t_{s,1}}  + \dfrac{1}{1+\Omega t_{s,2}} \right)^{1/2}&\textrm{~for~} t_{s,1}\gg \Omega^{-1},
\end{cases}
\end{equation}
where $\delta v_\mathrm{g} = \alpha^{1/2} c_\mathrm{s}$ is the mean random velocity of the largest turbulent eddies, and $t_\eta= \mathrm{Re_t}^{1/2} t_L$ is the turn-over time of the smallest eddies. The turbulence Reynolds number $\mathrm{Re_t}= \alpha c_\mathrm{s}^2 / (\Omega \nu_\mathrm{mol} )$, with the molecular viscosity $\nu_\mathrm{mol}=v_\mathrm{th} \lambda_\mathrm{mfp}/2$.

The relative velocity from radial drift just equals $\Delta v_r = |v_\mathrm{drift}(\Omega t_{s,1}) - v_\mathrm{drift}(\Omega t_{s,2})|$, with the drift velocity given by Eq. \ref{eq:v_drift}. The azimuthal relative velocity is obtained in a similar way, as $\Delta v_\phi = |v_\phi (\Omega t_{s,1}) - v_\phi(\Omega t_{s,2})|$ with
\begin{equation}
v_\phi = - \frac{\eta v_\mathrm{K}}{1+(\Omega t_s)^2}.
\end{equation}
Finally, the total relative velocity is given by
\begin{equation}
v_\mathrm{rel} = \sqrt{ (\Delta v_\mathrm{BM})^2 + (\Delta v_\mathrm{turb})^2 + (\Delta v_r)^2 + (\Delta v_\phi)^2}.
\end{equation}
The particle stopping times and relative velocities are calculated in the midplane of the gas disk, as this is where most of the coagulation occurs.

\section{Particle porosity evolution}\label{sec:phi}
Initially, particles grow through low-energy hit-and-stick collisions. During this growth phase, the fractal dimension is ${\sim}2$ \citep{kempf1999}, and the porosity is given by 
\begin{equation}\label{eq:fractal}
\phi \simeq (m / m_\bullet)^{1/2},
\end{equation}
where $m_\bullet$ is the monomer mass. The fractal growth regime ends when collisions become energetic enough for compaction \citep{dominiktielens1997} or when gas ram pressure compaction becomes effective \citep{kataoka2013a}. 

For low internal densities, \citet{kataoka2013a} found that the external pressure a dust aggregate can just withstand equals
\begin{equation}\label{eq:P_c}
P_\mathrm{crit} = \frac{E_{\mathrm{roll}}}{a_\bullet^3}\phi^3.
\end{equation}
An important parameter is the critical rolling energy $E_\mathrm{roll}$ \citep{dominiktielens1997}. Based on experimental investigations \citep{heim1999, gundlach2011} and theoretical work \citep{krijt2014} we obtain 
\begin{equation}
E_\mathrm{roll} = \left(\dfrac{a_\bullet}{\mathrm{1~\mu m}} \right)^{5/3} \times \begin{cases} 
~1.8\times10^{-7} \mathrm{~erg}  & \textrm{~~for water ice},  \vspace{3mm} \\
~8.5\times10^{-9} \mathrm{~erg}  & \textrm{~~for silicate}.
\end{cases}
\end{equation}
The pressure of Eq. \ref{eq:P_c} can then be compared to the pressure arising form the surrounding gas and from self-gravity \citep{kataoka2013c}
\begin{equation}\label{eq:P_gg}
P_\mathrm{gas} = \frac{v_\mathrm{dg} m}{\pi a^2 t_s}, \\
P_\mathrm{grav} = \frac{Gm^2}{\pi a^4},
\end{equation}
with $G$ the gravitational constant. 

In our semi-analytical model, the porosity is determined by fractal growth followed by gas- and eventually self-gravity compaction. Specifically, we make use of Eq. \ref{eq:fractal}, until the external pressures become too large for the aggregate to withstand. At that point, we obtain the porosity by setting Eq. \ref{eq:P_c} equal to Eq. \ref{eq:P_gg}. Of these processes, only gas pressure compaction varies with disk location, being more efficient in at smaller disk radii. 

\begin{figure}
\centering
\includegraphics[clip=,width=.95\linewidth]{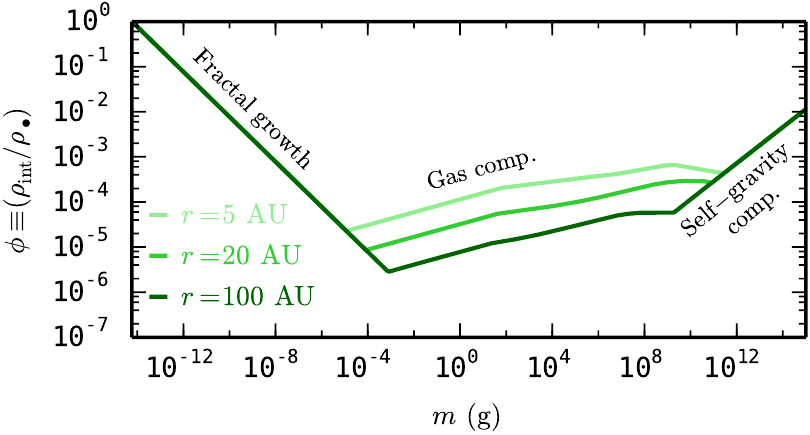}
\caption{Particle porosity as a function of aggregate mass at various locations in the benchmark disk model (see Table \ref{tab:benchmark}), assuming $0.1\mathrm{~\mu m}$ monomers. When calculating $\phi(m,r)$, we assume the aggregates grow though hit-and-stick collisions, followed by gas- and self-gravity compaction (see text). Gas compaction is more efficient at smaller radii.}
\label{fig:porosity}
\end{figure}

As particles only move inward in the cases considered in this work (Fig. \ref{fig:panoptic}), we can construct a single function $\phi(m,r)$ that is independent of the dust particle's growth- and drift history. More specifically, for our assumptions (i.e., fractal growth followed by gas- and self-gravity compaction), $\phi(m,r_1) \geq \phi(m,r_2 > r_1)$. Figure \ref{fig:porosity} shows porosity as a function of mass for three different radii, assuming $0.1\mathrm{~\mu m}$ monomers and the disk properties of Table \ref{tab:benchmark}. 

\end{appendix}

\end{document}